\documentclass[conference]{IEEEtran}
\IEEEoverridecommandlockouts
\usepackage{cite}
\usepackage{amsmath,amssymb,amsfonts}
\usepackage{algorithmic}
\usepackage{graphicx}
\usepackage{textcomp}
\usepackage{xcolor}

\usepackage{booktabs}
\usepackage{xcolor}
\usepackage{authblk}
\usepackage{multirow}

\usepackage[normalem]{ulem}
\useunder{\uline}{\ul}{}

\def\BibTeX{{\rm B\kern-.05em{\sc i\kern-.025em b}\kern-.08em
    T\kern-.1667em\lower.7ex\hbox{E}\kern-.125emX}}

\newcommand{\framework}{\textit{Dynamite}}

\title{DYNAMITE: Dynamic Defense Selection for Enhancing Machine Learning-based Intrusion Detection Against Adversarial Attacks}

\author[]{Jing Chen\textsuperscript{*}, Onat Gungor\textsuperscript{*}, Zhengli Shang, Elvin Li, Tajana Rosing}
\affil[]{Department of Computer Science and Engineering, University of California, San Diego}
\affil{\textit{\{jic128, ogungor, z4shang, ell009, tajana\}@ucsd.edu}}

\begin{document}
\maketitle

\begingroup
\renewcommand\thefootnote{\textasteriskcentered}
\footnotetext{Both authors contributed equally to this research.}
\endgroup


\begin{abstract}
The rapid proliferation of the Internet of Things (IoT) has introduced substantial security vulnerabilities, highlighting the need for robust Intrusion Detection Systems (IDS). Machine learning-based intrusion detection systems (ML-IDS) have significantly improved threat detection capabilities; however, they remain highly susceptible to adversarial attacks. While numerous defense mechanisms have been proposed to enhance ML-IDS resilience, a systematic approach for selecting the most effective defense against a specific adversarial attack remains absent. To address this challenge, we propose \framework{}, a dynamic defense selection framework that enhances ML-IDS by intelligently identifying and deploying the most suitable defense using a machine learning-driven selection mechanism. Our results demonstrate that \framework{} achieves a 96.2\% reduction in computational time compared to the Oracle, significantly decreasing computational overhead while preserving strong prediction performance. \framework{} also demonstrates an average F1-score improvement of 76.7\% over random defense and 65.8\% over the best static state-of-the-art defense.


\end{abstract}


\section{Introduction}\label{sec:intro}   
The Internet of Things (IoT) systems connect numerous devices that communicate and share data, enabling smart applications in sectors like healthcare, manufacturing, and transportation \cite{zarpelao2017survey}. IoT systems are particularly susceptible to cyber threats due to their inter-connectivity, resource constraints, and diverse configurations \cite{abiodun2021review}. Consequently, ensuring robust security measures is essential to safeguard these systems against potential attacks. Intrusion Detection Systems (IDS) play a crucial role in identifying and responding to malicious activities within IoT networks by monitoring network traffic and system behavior \cite{zarpelao2017survey}. The integration of machine learning (ML) into IDS has significantly improved their effectiveness in detecting and mitigating cyber threats. ML-IDS possess the capability to analyze vast amounts of data, identify latent patterns, and detect cyberattacks that conventional methods may overlook \cite{da2019internet}. Thus, ML-IDS serve as a robust approach for enhancing IoT security by addressing evolving threats. However, the rise of adversarial attacks poses a significant challenge to the effectiveness of ML-IDS \cite{gungor2024rigorous}. These attacks allow malicious activities to go undetected and harm the security of IoT systems, leading to compromised operations, data breaches, and significant financial losses \cite{mishra2021internet}. 

\begin{figure}[]
\centering\includegraphics[scale=0.37]{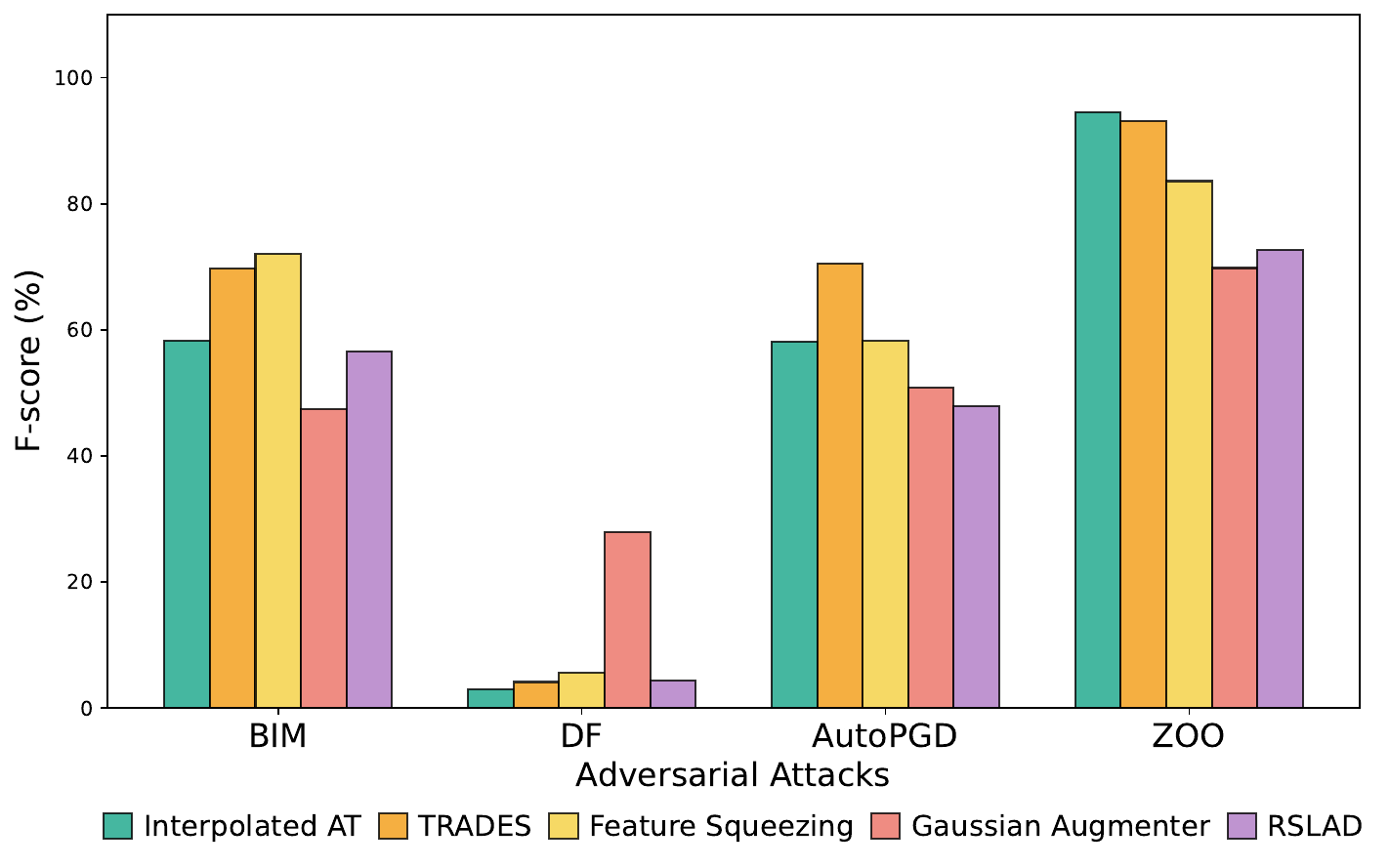}
\caption{SOTA Defense Performance Against Adversarial Attacks}
\label{fig-motivation}
\end{figure}

Developing effective defenses against adversarial attacks is crucial for maintaining the reliability and robustness of ML-IDS \cite{alotaibi2023adversarial}. Several strategies, both general and specific to ML-IDS, have been proposed, including adversarial training \cite{madry2017towards, zhang2019theoretically}, modifications to the training process \cite{papernot2016distillation, zi2021revisiting}, input transformation techniques \cite{xu2017feature}, and methods for adversarial attack detection \cite{debicha2023adv, debicha2023tad}. However, the effectiveness of defense mechanisms varies depending on the specific type of attack they are intended to mitigate \cite{wang2021adversarial}. Given that adversarial attacks can differ in their techniques and objectives, tailored defense strategies are necessary to effectively address each distinct scenario. Fig. \ref{fig-motivation} demonstrates that no single defense model (represented by different colors) is universally effective against all adversarial attacks, with the optimal defense varying depending on the specific nature of the attack (as shown on the x-axis). This variability highlights the limitation of relying on a singular defense mechanism for comprehensive protection. It further emphasizes the importance of a dynamic defense selection mechanism that adaptively assigns the most appropriate defense for each attack scenario. Such an approach is crucial for achieving robust security, as it ensures the real-time deployment of the most effective defense in response to the evolving nature of adversarial attacks.

We propose an adaptive ML-IDS defense framework that ensures robust protection by dynamically selecting the most suitable defense for each adversarial attack. In contrast to traditional approaches that rely on static defenses, our framework adaptively mitigates the impact of these attacks. As depicted in Fig. \ref{framework}, our framework, \framework{}, follows a comprehensive pipeline, starting with data preprocessing to clean, normalize, and encode features. The processed data is used to train both a baseline model and several defense models for robust evaluation. Adversarial samples are then generated using various attack strategies with different intensities to simulate real-world scenarios. Defense models are assessed based on their performance against these adversarial samples, and performance metrics are recorded to label the samples with their most effective defense. Finally, an ML classifier is trained on this labeled data to dynamically predict the most suitable defense model to unseen adversarial attacks. Our experiments on different intrusion datasets demonstrate that \framework{} outperforms both random defense and the best static defense, yielding an average F1-score improvement of 76.7\% and 65.8\%, respectively. Additionally, \framework{} significantly enhances computational efficiency, achieving a 96.2\% computational time reduction over the Oracle with only a 1.7\% F1-score gap. 
These results underscore the effectiveness of \framework{} as a scalable and efficient defense strategy for intrusion detection, successfully balancing high accuracy with reduced computational overhead.



\section{Related Work}\label{sec:related}

The growing dependence on computer networks and the expansion of the Internet of Things (IoT) have introduced significant security challenges, driven by the increasing complexity and diversity of these interconnected systems. Intrusion Detection Systems (IDS) are designed to monitor network activity and detect malicious behavior. The integration of machine learning (ML) has enhanced their capability to identify complex and evolving attack patterns with greater accuracy. However, ML-based IDS are vulnerable to adversarial attacks, where carefully crafted input perturbations deceive the models into making incorrect predictions \cite{gungor2024roldef}. To enhance resilience against adversarial attacks, various defense strategies have been proposed, which can be broadly categorized into adversarial training \cite{madry2017towards, shafahi2019adversarial, zhang2019theoretically}, modifying the training process \cite{papernot2016distillation, zi2021revisiting}, and using supplementary networks \cite{xu2017feature, kassam2012signal}. Several efforts have been directed toward developing adversarial defense mechanisms specifically tailored to enhance the robustness of ML-IDS against adversarial attacks. Han et al. \cite{han2021evaluating} address traffic-space attacks targeting ML-based NIDS, proposing a defense scheme that reduces evasion rates across multiple attack scenarios. Debicha et al. \cite{debicha2023adv} introduce Adv-Bot, a framework for generating adversarial botnet traffic to test and strengthen IDS defenses. Additionally, Debicha et al. \cite{debicha2023tad} present a transfer learning-based framework that employs multiple adversarial detectors to improve detection rates.   Existing studies on ML-based IDS defenses against adversarial attacks often focus on isolated mechanisms or manual selection, limiting their generalizability.

In contrast, our framework incorporates multiple SOTA defenses and dynamically selects the most effective one based on the performance across the dataset. Rather than requiring extensive manual tuning or being restricted to specific attack types, our approach generalizes to diverse adversarial scenarios by training a classifier to predict the most suitable defense. Once deployed, the algorithm processes each new sample individually and predicts the optimal defense in real-time based on learned patterns. This shift from static to performance-based dynamic defense selection enables our framework to offer robust protection across a broader range of adversarial threats, distinguishing it from existing methods \cite{madry2017towards, shafahi2019adversarial, zhang2019theoretically, papernot2016distillation, zi2021revisiting, xu2017feature, kassam2012signal}.

\section{Proposed Framework}\label{sec:framework}

\begin{figure*}[]
      \centering
      \includegraphics[scale=0.85]{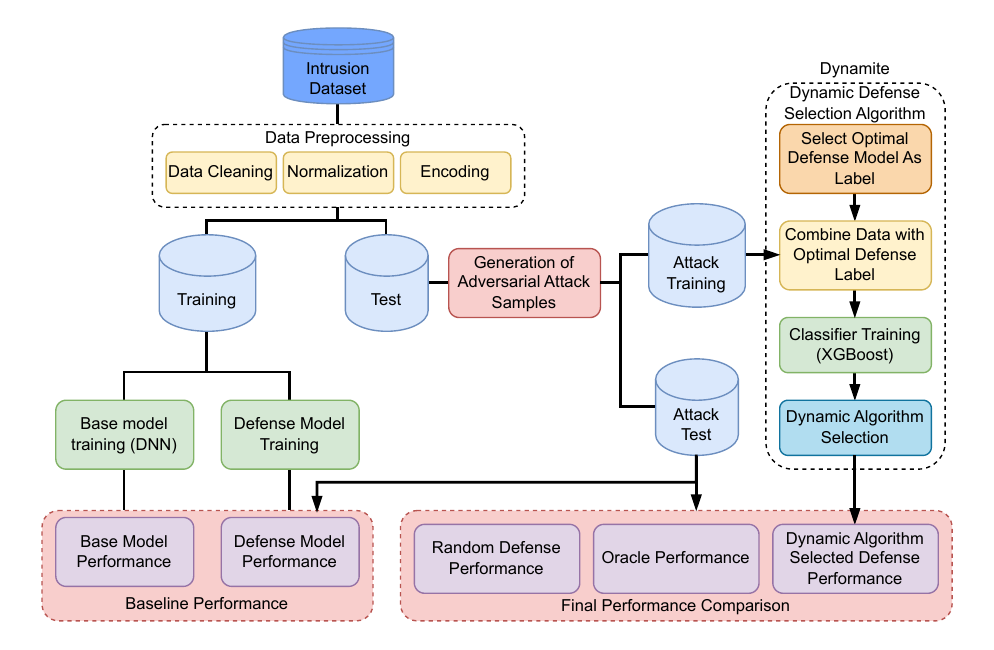}
      \caption{\framework{} is a dynamic defense selection framework designed to enhance ML-based intrusion detection against adversarial attacks. It integrates baseline model training, adversarial sample generation, defense model training, and dynamic defense assignment to effectively address a wide range of attack scenarios.}
      \label{framework}
\end{figure*}

We propose \framework{}, a dynamic defense selection framework designed to strengthen ML-based IDS against adversarial attacks. By addressing the challenges posed by diverse attack types and varying intensities, \framework{} provides a robust defense solution. Our framework integrates key components, including adversarial sample generation, defense model training, and dynamic defense selection, forming a comprehensive pipeline to evaluate and mitigate adversarial threats effectively. As shown in Figure \ref{framework}, the process begins with data preprocessing, where raw data is cleaned, normalized, and encoded. This preprocessed data is then used to train both a baseline DNN model and various defense models. To simulate real-world scenarios, adversarial datasets are generated using multiple attack models, providing a comprehensive benchmark for testing the framework’s effectiveness. Next, we evaluate and record the performance of several state-of-the-art defense models to identify the most effective strategies for different adversarial scenarios. To enable dynamic defense selection, an XGBoost-based classifier analyzes patterns in the dataset to predict the most suitable defense model for ''new'' (unseen) adversarial samples. By integrating optimal defense labels derived from the performance matrix—which evaluates the effectiveness of each defense model against various adversarial attacks—the \framework{} dynamically selects the most suitable defense for each attack scenario. Finally, the \framework{}’s performance is compared to that of Oracle, the best static defense models, and random defense selection.


\subsection{Data Preprocessing}
This module involves data cleaning to remove redundant or irrelevant features, feature standardization to normalize numerical features for consistent scaling, and categorical encoding to convert classification features into numerical representations compatible with ML models. After preprocessing the data, it is divided into training and test sets. The training set is used for baseline model and defense model training, while the test set is reserved for adversarial attack generation and final evaluation.


\subsection{Generation of Adversarial Attack Samples}
\subsubsection{Selected Adversarial Attacks} We employ six widely used adversarial attacks: BIM\cite{kurakin2018adversarial}, FGSM\cite{goodfellow2014explaining}, PGD\cite{madry2017towards}, DF\cite{moosavi2016deepfool}, AutoPGD\cite{croce2020reliable}, and ZOO\cite{chen2017zoo}. These attacks use gradient-based and query-based methods to generate adversarial samples, introducing input perturbations to manipulate model predictions. The perturbation amount, controlled by the epsilon ($\varepsilon$) value, determines the intensity of the attacks, ranging from subtle to more pronounced alterations. This setup enables comprehensive testing under diverse and realistic conditions, offering valuable insights into the framework’s performance against various adversarial scenarios. 

\subsubsection{Adversarial Dataset Generation}
The generation process involves applying each attack model to the dataset, with epsilon ($\varepsilon$) values adjusted to simulate varying levels of adversarial intensity. A unique adversarial dataset is generated for each combination of six attack methods (BIM, FGSM, PGD, DF, AutoPGD, and ZOO) and four epsilon values (0.01, 0.1, 0.2, 0.3), resulting in a total of 24 distinct datasets. Each attack is applied to the test dataset, maintaining the same sample size as the original. This ensures consistent evaluation while introducing adversarial perturbations based on attack type and intensity. After generating adversarial attack samples, we split them into two sets: attack training and attack test. The training portion is used to train our dynamic defense selection model, while the test portion is used for final evaluation.  

\subsection{Baseline Model Training}
To establish a performance baseline, a Deep Neural Network (DNN) \cite{al2021x} is trained on the original, unperturbed dataset. The model is then evaluated under different adversarial attack configurations, providing a reference for assessing the effectiveness of defense strategies. This baseline serves as a crucial benchmark, illustrating the impact of adversarial attacks on model performance and emphasizing the importance of robust defense mechanisms and dynamic selection approaches.


\subsection{Defense Model Training}

We evaluate the effectiveness of nine state-of-the-art defenses against adversarial attacks: Projected Gradient Descent Adversarial Training \cite{madry2017towards}, Interpolated Adversarial Training \cite{lamb2019interpolated}, Tradeoff-inspired Adversarial Defense via Surrogate-loss Minimization (TRADES) \cite{zhang2019theoretically}, Free Adversarial Training \cite{shafahi2019adversarial}, Gaussian Augmenter \cite{zantedeschi2017efficient}, Defensive Distillation \cite{papernot2016distillation}, Robust Soft Label Adversarial Distillation (RSLAD) \cite{zi2021revisiting}, Feature Squeezing \cite{xu2017feature}, and Gaussian Noise \cite{kassam2012signal}. These defenses were selected for their diverse mechanisms, including adversarial training, data augmentation, loss optimization, and input preprocessing. This selection ensures a comprehensive and diverse evaluation of defense strategies across multiple methodologies. To address varying defense requirements, we introduce multiple parameter configurations for certain models. For RSLAD, configurations like RSLAD10 and RSLAD100 adjust optimization strength to evaluate robustness tradeoffs. 
This approach systematically assesses adaptability to different adversarial perturbation levels. Applying these defense methods to diverse adversarial datasets enables the framework to evaluate model adaptability and performance across attack scenarios. These defenses form the basis of the dynamic selection mechanism, allowing the framework to deploy the most effective strategy for each adversarial sample, ensuring robust performance under varying attack types and intensities.

\subsection{Optimal Defense Identification}

\subsubsection{Constructing Attack Training and Attack Test Data}
The attack training and attack test data are created using a subset of the 24 adversarial datasets—generated using different attack methods and epsilon values from the test set—ensuring a distinction between known and unknown data during model evaluation. Specifically, the datasets with an epsilon value of 0.1 (8 datasets) are used as attack training data, representing the known data. The remaining datasets, with other epsilon values (16 datasets), serve as attack test data, representing the unknown data. This setup allows the framework to assess its ability to generalize beyond the perturbation strengths encountered during training. 

\subsubsection{Optimal Defense Selection}
To assess the defense models, we process attack training data through all nine defenses and record key metrics, such as the macro F1-score. This generates a performance matrix, where each entry represents a defense model's effectiveness against a specific adversarial dataset. The matrix serves as a basis for comparing defenses and identifying best strategies, offering insights into how each model addresses adversarial perturbations. To determine the most effective defense for each adversarial sample, we analyze the performance metrics of all nine defense models and select the highest-performing defense for each sample. This selected model is then used as the label, which forms the ground truth for training our dynamic defense selection mechanism. 

\subsection{Dynamic Defense Selection Algorithm}
Our dynamic defense selection algorithm is designed to adaptively assign the most suitable defense model to each adversarial sample, helping to maintain strong performance across varying attack conditions. To achieve this, we utilize XGBoost (XGB) \cite{chen2016xgboost} as a classifier. During training, we combine attack training data with their corresponding optimal defense labels and feed them into XGB. This enables the model to learn the relationships between network features and the most effective defense models. Through this process, XGB identifies key features and establishes a mapping between attack patterns and defense strategies. In the classification phase, XGB can efficiently process the attack test data and assign the most appropriate defense model to each adversarial sample. Furthermore, it generalizes to new adversarial data, dynamically adapting to different attack types and intensities in real-time. This dynamic adaptation mechanism eliminates the reliance on static defense, allowing the framework to adjust its defense strategies based on the unique characteristics of each attack. By dynamically selecting defenses, the framework enhances its resilience against unseen adversarial threats, ensuring robust and consistent performance across diverse attack scenarios.

\subsection{Final Performance Comparison}

During the final evaluation phase, each sample is assigned to a corresponding defense model (e.g., TRADES, RSLAD), ensuring that the selected strategy effectively mitigates the adversarial attack. 
To comprehensively assess \framework{}'s effectiveness, we compute the Macro F1-Score for each selected defense model, offering a holistic measure of overall performance. This final metric is then compared with other baseline approaches, such as Oracle, random defense, and the best static defense, highlighting \framework{}'s robustness and adaptability across diverse adversarial scenarios.

\begin{table*}[]
\centering
\caption{Final Performance (Macro F1-score) Comparison}
\begin{tabular}{|c|ccccc|ccccc|}
\hline
        & \multicolumn{5}{c|}{UNSW-NB15}                                                                                                                       & \multicolumn{5}{c|}{WUSTL-IIoT}                                                                                                                      \\ \hline
(\%)    & \multicolumn{1}{c|}{No Defense} & \multicolumn{1}{c|}{Dynamite}    & \multicolumn{1}{c|}{Oracle}         & \multicolumn{1}{c|}{Random} & Best-Static\cite{madry2017towards} & \multicolumn{1}{c|}{No Defense} & \multicolumn{1}{c|}{Dynamite}    & \multicolumn{1}{c|}{Oracle}         & \multicolumn{1}{c|}{Random} & Best-Static\cite{xu2017feature} \\ \hline
BIM     & \multicolumn{1}{c|}{30.78}      & \multicolumn{1}{c|}{81.17}       & \multicolumn{1}{c|}{81.78}          & \multicolumn{1}{c|}{64.64}  & 68.73       & \multicolumn{1}{c|}{28.44}      & \multicolumn{1}{c|}{77.77}       & \multicolumn{1}{c|}{87.16}          & \multicolumn{1}{c|}{53.06}  & 72.08       \\ \hline
FGSM    & \multicolumn{1}{c|}{43.20}      & \multicolumn{1}{c|}{81.22}       & \multicolumn{1}{c|}{83.23}          & \multicolumn{1}{c|}{66.64}  & 81.99       & \multicolumn{1}{c|}{29.51}      & \multicolumn{1}{c|}{71.75}       & \multicolumn{1}{c|}{77.73}          & \multicolumn{1}{c|}{54.28}  & 59.57       \\ \hline
PGD     & \multicolumn{1}{c|}{30.78}      & \multicolumn{1}{c|}{81.17}       & \multicolumn{1}{c|}{81.86}          & \multicolumn{1}{c|}{64.64}  & 68.73       & \multicolumn{1}{c|}{28.44}      & \multicolumn{1}{c|}{77.78}       & \multicolumn{1}{c|}{87.16}          & \multicolumn{1}{c|}{53.06}  & 72.08       \\ \hline
DF      & \multicolumn{1}{c|}{10.81}      & \multicolumn{1}{c|}{50.64}       & \multicolumn{1}{c|}{55.19}          & \multicolumn{1}{c|}{20.56}  & 39.69       & \multicolumn{1}{c|}{1.99}       & \multicolumn{1}{c|}{25.37}       & \multicolumn{1}{c|}{27.91}          & \multicolumn{1}{c|}{6.15}   & 5.61        \\ \hline
AutoPGD & \multicolumn{1}{c|}{29.52}      & \multicolumn{1}{c|}{80.15}       & \multicolumn{1}{c|}{82.39}          & \multicolumn{1}{c|}{65.97}  & 71.03       & \multicolumn{1}{c|}{26.03}      & \multicolumn{1}{c|}{56.52}       & \multicolumn{1}{c|}{76.83}          & \multicolumn{1}{c|}{50.91}  & 58.20       \\ \hline
ZOO     & \multicolumn{1}{c|}{83.49}      & \multicolumn{1}{c|}{90.57}       & \multicolumn{1}{c|}{90.77}          & \multicolumn{1}{c|}{87.15}  & 87.36       & \multicolumn{1}{c|}{79.24}      & \multicolumn{1}{c|}{91.18}       & \multicolumn{1}{c|}{94.61}          & \multicolumn{1}{c|}{81.86}  & 83.59       \\ \hline
Average & \multicolumn{1}{c|}{38.10}      & \multicolumn{1}{c|}{{\ul 77.49}} & \multicolumn{1}{c|}{\textbf{79.20}} & \multicolumn{1}{c|}{61.60}  & 69.59       & \multicolumn{1}{c|}{32.28}      & \multicolumn{1}{c|}{{\ul 66.73}} & \multicolumn{1}{c|}{\textbf{75.23}} & \multicolumn{1}{c|}{49.89}  & 58.52       \\ \hline
\end{tabular}
\label{model_Performance}
\end{table*}

\section{Experimental Analysis}\label{sec:setup}

\subsection{Baselines}

\textbf{No Defense:} This baseline evaluates the performance of a standard DNN model under adversarial attacks without defenses, establishing the lower performance bound.

\textbf{Random Defense:} The random defense performance is assessed by randomly selecting a defense model for each adversarial dataset 100 times. The average performance is then computed, providing a benchmark for evaluating a non-deterministic, uninformed defense selection strategy.

\textbf{Best Static Defense:} The best static defense evaluates defense models on attack training data to identify the most effective defense model. The model with the highest average performance across all adversarial attack types is selected and tested on the attack test data to assess its effectiveness.

\textbf{Oracle Defense:} The Oracle represents the theoretical upper bound of defense performance, derived by selecting the best-performing defense for each of the 24 adversarial datasets across all defense models. Comparing our framework's performance to the Oracle shows how closely the dynamic defense mechanism approximates this ideal.

\subsection{Selected Datasets}

\textbf{WUSTL-IIoT}\cite{zolanvari2021wustl}: The WUSTL-IIoT dataset, designed for cybersecurity research in Industrial Internet of Things (IIoT) environments, replicates real-world industrial operations for realistic cyber-attack simulations. It includes network traffic data from IIoT testbeds across various attack scenarios, with 41 features and 1M samples, supporting the development of ML-driven security solutions for industrial settings.

\textbf{UNSW-NB15}\cite{unsw15}: The UNSW-NB15 dataset is designed for network intrusion detection, combining real-world normal network activities with synthetically generated attack behaviors. It includes nine attack types and features extracted using both traditional and novel techniques. With 43 features and 278K samples, it serves as a key benchmark for developing and evaluating intrusion detection systems.

\subsection{Experimental Setup}

\textbf{Hardware}: We conduct our experiments on a Linux virtual machine server equipped with a 16-core CPU, 32 GB of RAM, and an NVIDIA A100 GPU with 80 GB of memory.

\textbf{Evaluation Metric}: We select the Macro F1 score as our evaluation metric because it offers a balanced assessment of model performance across all classes, independent of class distribution. This metric is especially pertinent for datasets with imbalanced attack types, as it ensures that minority classes are appropriately represented in the evaluation.








\textbf{\framework{} Performance Scoring:} The final defense performance is evaluated using a weighted scoring formula, which combines the number of samples handled by each defense model and its performance:

\begin{equation}
\text{Score} = \sum_{i=1}^{N} \left( \frac{\text{Sample Count}_i}{\text{Total Samples}} \times \text{Model Performance}_i \right)  
\end{equation}

where \( \text{Sample Count}_i \) represents the number of adversarial samples assigned to the \(i\)-th defense model, and \( \text{Model Performance}_i \) denotes the Macro F1 score of the \(i\)-th defense model. This formula calculates a weighted average of the defense models' performances, with the weight determined by the proportion of samples managed by each model. By doing so, it provides a holistic view of how well the dynamic algorithm selection performs across all assigned samples. A higher score reflects both the framework’s ability to assign the most suitable defense models and the overall effectiveness of those models in mitigating adversarial impacts.

\section{Results}\label{sec:results}

\subsection{\framework{} Defense Performance}
Table \ref{model_Performance} provides a comparative analysis of \framework{} against selected baselines (no defense, oracle, random, and best static), emphasizing key performance variations across different attacks and datasets. The table reports the macro F1 score for each adversarial attack, as well as the average scores across all attack scenarios. It is evident that \framework{} substantially outperforms the best static, random, and no defense baselines, achieving performance nearly equivalent to that of the Oracle. These results underscore Dynamite's capability in improving model robustness while reducing adversarial impact.

\begin{figure}[]
\centering\includegraphics[scale=0.35]{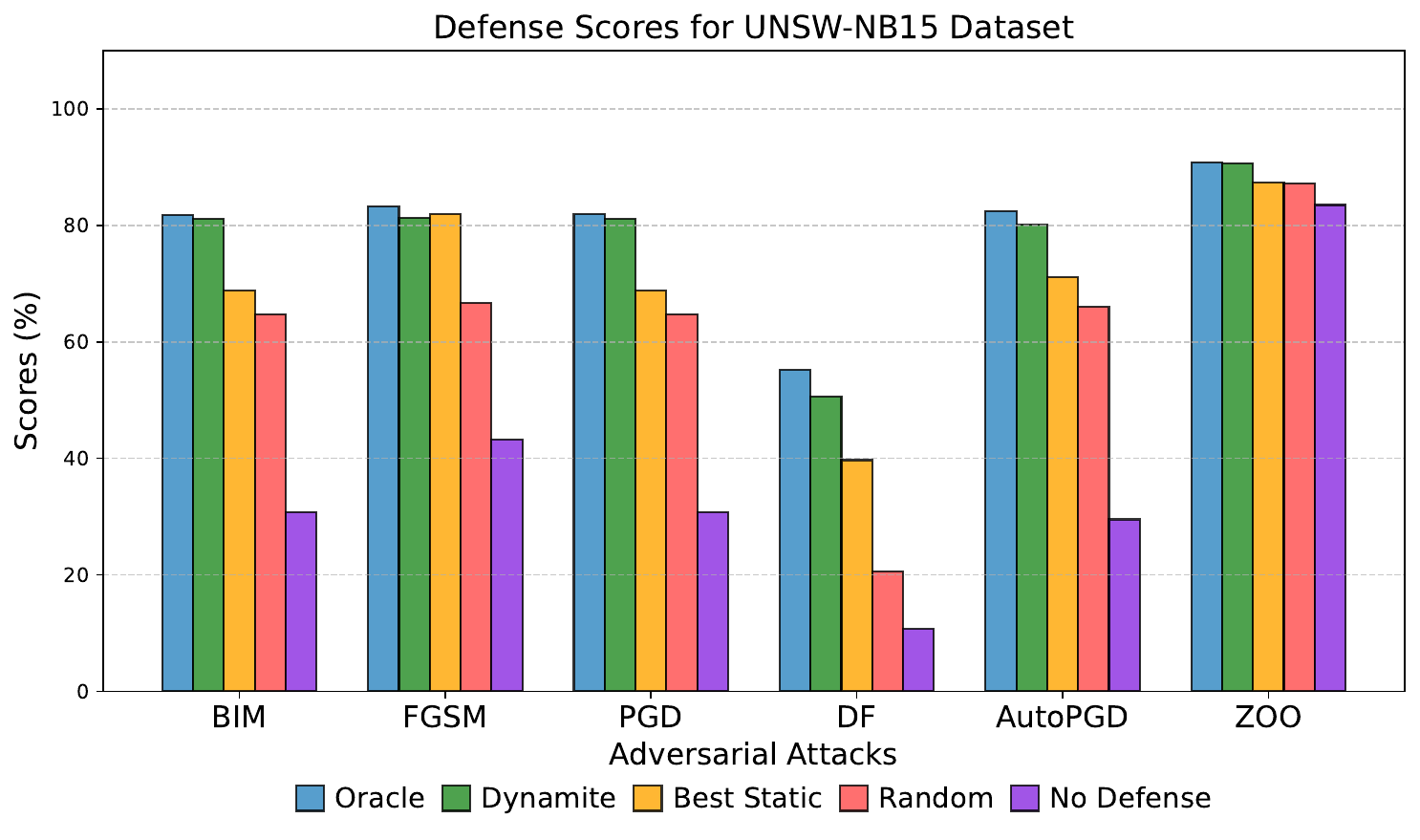}
\caption{Prediction Performance Comparison (UNSW-NB15)}
\label{UNSW_Performance}
\end{figure}

\begin{figure}[]
\centering\includegraphics[scale=0.35]{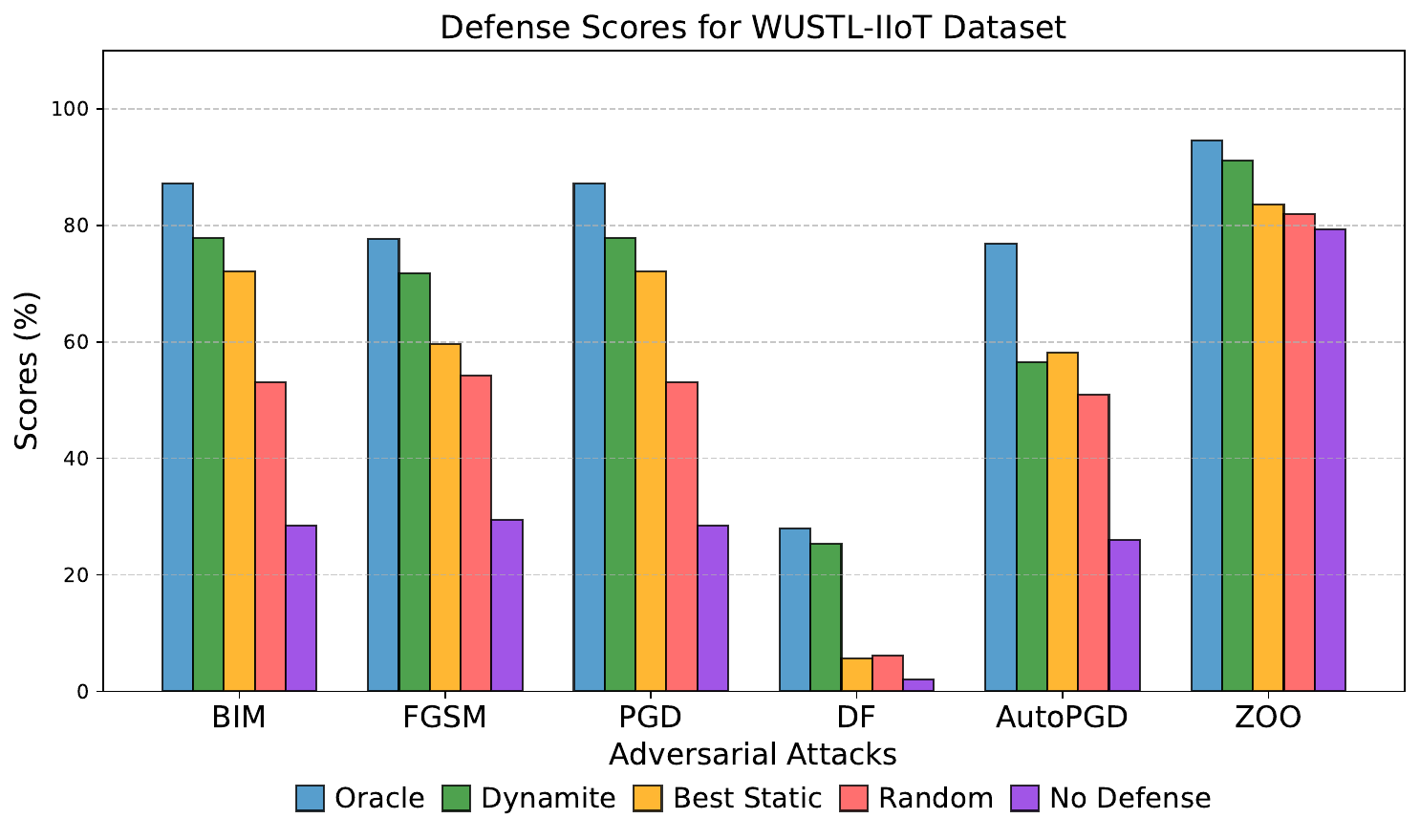}
\caption{Prediction Performance Comparison (WUSTL-IIoT)}
\label{WUSTL_Performance}
\end{figure}

\subsubsection{Comparison with Random and Best Static Defenses}
\framework{} exhibits the most significant improvement in performance for the DF attack among all considered adversarial attacks. \framework{} achieves substantial improvements over both random and best static defenses, with performance gains of 146.3\% and 27.6\% on the UNSW-NB15 dataset, and 312.5\% and 352.2\% on the WUSTL-IIoT dataset, respectively, for the DF attack. This demonstrates that, especially in the case of stronger attacks, \framework{} substantially outperforms random and best static, highlighting its effectiveness and reliability in optimizing defense model allocation. As shown in Fig. \ref{UNSW_Performance} and Fig. \ref{WUSTL_Performance}, static defenses are more vulnerable to white-box attacks, as attackers can exploit their weaknesses, whereas in black-box attacks like ZOO, static defenses maintain relatively stable performance. However, \framework{} still achieves a notable performance gain, improving up to 14\% over the best static defense. Table \ref{improvement_rate} presents both the maximum and average performance improvements of \framework{} compared to the random and best static defenses. Our method also shows average improvements over both baselines, highlighting its effectiveness and adaptability in dynamically assigning optimal defense strategies.

\subsubsection{Comparison with Oracle}

\framework{} demonstrates a remarkably small performance gap with the Oracle, underscoring its effectiveness in dynamically selecting defenses across diverse adversarial conditions. In the UNSW dataset, \framework{} achieves 90.57\% on the ZOO attack, with only a 0.2\% gap from the Oracle’s 90.77\% in a less challenging scenario. Even in the most challenging case, the DF attack, the gap remains limited to just 4.55\%. Similarly, in the WUSTL dataset, \framework{} reduces the gap to only 2.54\% for the DF attack. These results highlight \framework{}'s ability to allocate defenses effectively without requiring exhaustive model evaluations, making it both efficient and overhead. On average, the performance difference remains minimal, with a 1.71\% gap in UNSW and 8.5\% in WUSTL, reinforcing \framework{}’s adaptability even in more complex adversarial scenarios. 

\textbf{Insights:} The results reflect \framework{}’s ability to approach the theoretical upper bound established by the Oracle while demonstrating its flexibility in handling diverse adversarial scenarios. Since the Oracle determines the best performance by testing every dataset across all models, achieving this level of optimality in practice would require significant effort and resources, making it impractical for real-world applications. In contrast, the \framework{} dynamically allocates defense strategies using a machine learning-based approach. This method eliminates the need for manually selecting the optimal defense for each attack, showcasing the \framework{}’s adaptability in addressing complex adversarial scenarios. Moreover, the \framework{} offers enhanced scalability in practical applications, achieving performance levels close to the Oracle without requiring explicit identification of each attack type.



\begin{table}[]
\centering
\caption{\framework{} F1-score Improvement Rate}
\begin{tabular}{|c|cc|cc|}
\hline
\multirow{2}{*}{\begin{tabular}[c]{@{}c@{}}Improvement\\ Rate (\%)\end{tabular}} & \multicolumn{2}{c|}{UNSW-NB15} & \multicolumn{2}{c|}{WUSTL-IIoT} \\ \cline{2-5} 
 & \multicolumn{1}{c|}{Random} & Best-Static & \multicolumn{1}{c|}{Random} & Best-Static \\ \hline
Max & \multicolumn{1}{c|}{146.3} & 27.6 & \multicolumn{1}{c|}{312.5} & 352.2 \\ \hline
Average & \multicolumn{1}{c|}{40.8} & 13.2 & \multicolumn{1}{c|}{76.7} & 65.8 \\ \hline
\end{tabular}
\label{improvement_rate}
\end{table}

\subsection{Overhead Analysis}

The overhead analysis examines the computational efficiency of \framework{} by comparing the time required for defense selection per sample with that of the Oracle and Best-static. Table \ref{processing_time} presents the processing time in ms/sample, highlighting the substantial difference between the two algorithms. For the UNSW-NB15 dataset, the Oracle requires 22.08 ms/sample, whereas the \framework{} reduces this to 0.84 ms/sample, achieving a time reduction of 96.2\%. Similarly, in the WUSTL-IIoT dataset, the Oracle requires 24.28 ms/sample, while the \framework{} reduces this to 0.67 ms/sample, resulting in a 97.3\% computational time reduction. These results demonstrate that \framework{} substantially reduces computational overhead in comparison to the Oracle. When evaluated alongside Best-static, which requires 2.20 ms/sample in UNSW-NB15 and 2.21 ms/sample in WUSTL-IIoT, \framework{} further reduces processing time by 61.8\% and 69.8\%, respectively. While Best-static is more efficient than the Oracle, it still lacks the adaptability of \framework{}. This contrast underscores \framework{}'s efficiency in accelerating defense selection while maintaining strong performance, making it a viable adversarial defense solution. 

\begin{table}[]
\centering
\renewcommand{\arraystretch}{1.2}
\setlength{\tabcolsep}{8pt}
\caption{Processing time per sample comparison}
\begin{tabular}{lcc}
\toprule
\textbf{ms/Sample} & \textbf{UNSW-NB15} & \textbf{WUSTL-IIoT} \\
\midrule
Dynamite & 0.8396 & 0.6670 \\
Oracle   & 22.077 & 24.276 \\
Best-static   & 2.198 & 2.211 \\
\bottomrule
\end{tabular}
\label{processing_time}
\end{table}



\section{Conclusion}\label{sec:conclusion}

Ensuring robust cybersecurity in machine learning-based intrusion detection remains a critical challenge due to its susceptibility to adversarial attacks. Although various defense mechanisms have been proposed for resilient ML-IDS, a systematic methodology for selecting the most effective defense tailored to specific adversarial attacks is still lacking. This paper proposes a dynamic framework that overcomes the limitations of static defenses by integrating multiple defense models and selecting the most effective one for each attack scenario. Unlike traditional approaches that rely on fixed or manually chosen defenses, \framework{} continuously adapts to evolving threats, achieving superior performance over both random selection and the best static defense. \framework{} also reduces computational overhead by 96.2\% compared to the Oracle, significantly decreasing computational time while maintaining strong defensive capabilities, with only a 1.7\% average F1-score loss.

\section*{Acknowledgments}
This work has been funded in part by NSF, with award numbers \#1826967, \#1911095, \#2003279, \#2052809, \#2100237, \#2112167, \#2112665, and in part by PRISM and CoCoSys, centers in JUMP 2.0, an SRC program sponsored by DARPA.

\bibliographystyle{IEEEtran}
\bibliography{biblio}

\end{document}